\documentclass[headings=standardclasses]{scrartcl}

\usepackage{geometry}
\usepackage[utf8]{inputenc}
\usepackage[T1]{fontenc}
\usepackage[onehalfspacing]{setspace}
\usepackage{charter}
\usepackage[charter]{mathdesign}
\usepackage{microtype}
\usepackage{amsmath}
\usepackage{amsthm}
\usepackage{tabulary}
\usepackage{longtable}
\usepackage{booktabs} 
\usepackage{diagbox}
\usepackage{ragged2e}

\usepackage{siunitx}
\usepackage{threeparttable}
\usepackage{bm}
\usepackage[pdftex]{graphicx}
\usepackage{rotating}
\usepackage{pdfpages}
\usepackage{pdflscape}
\usepackage{color}
\usepackage{array}
\usepackage[font=bf]{caption}
\usepackage{subcaption}
\usepackage{pgfplots}
\usepackage{pgfplotstable}
\usepackage{float}
\usepackage[title]{appendix}
\pgfplotsset{compat=1.9}
\usepackage{url}
\usepackage[section]{placeins}
\usepackage{natbib}
\usepackage{hyperref}
\hypersetup{
	colorlinks   = true,    
	urlcolor     = blue,    
	linkcolor    = blue,    
	citecolor    = blue      
}

\usepackage[boxed]{algorithm2e}
\SetKwFunction{KwInit}{Initialisation}
\SetKwComment{Comment}{\% }{}

\makeatletter
\renewcommand*\env@matrix[1][c]{\hskip -\arraycolsep
  \let\@ifnextchar\new@ifnextchar
  \array{*\c@MaxMatrixCols #1}}
\makeatother

\usepackage{xcolor}
\usepackage{soul}

\usepackage{verbatim}
\usepackage{mathtools}
\usepackage{changepage}

\begin{document}
\thispagestyle{empty}
\begin{spacing}{1.2}
\begin{flushleft}
\Large \textbf{A Dynamic Choice Model with Heterogeneous Decision Rules: Application in Estimating the User Cost of Rail Crowding} \\


\vspace{\baselineskip}
\normalsize
3 July 2020 \\
\vspace{\baselineskip}
\textsc{Prateek Bansal} \\
prateek.bansal@imperial.ac.uk \\
\vspace{\baselineskip}
\textsc{Daniel H\"{o}rcher} \\
d.horcher@imperial.ac.uk \\
\vspace{\baselineskip}
\textsc{Daniel J. Graham} \\
d.j.graham@imperial.ac.uk \\
\vspace{\baselineskip}
Transport Strategy Centre, Department of Civil and Environmental Engineering\\
Imperial College London, UK \\
\vspace{\baselineskip}
\end{flushleft}
\end{spacing}

\newpage
\pagenumbering{arabic}

\section*{Abstract}
Crowding valuation of subway riders is an important input to various supply-side decisions of transit operators. The crowding cost perceived by a transit rider is generally estimated by capturing the trade-off that the rider makes between crowding and travel time while choosing a route. However, existing studies rely on static compensatory choice models and fail to account for inertia and the learning behaviour of riders. To address these challenges, we propose a new dynamic latent class model (DLCM) which (i) assigns riders to latent compensatory and inertia/habit classes based on different decision rules, (ii) enables transitions between these classes over time, and (iii) adopts instance-based learning theory to account for the learning behaviour of riders. We use the expectation-maximisation algorithm to estimate DLCM, and the most probable sequence of latent classes for each rider is retrieved using the Viterbi algorithm. The proposed DLCM can be applied in any choice context to capture the dynamics of decision rules used by a decision-maker. We demonstrate its practical advantages in estimating the crowding valuation of an Asian metro's riders. To calibrate the model, we recover the daily route preferences and in-vehicle crowding experiences of regular metro riders using a two-month-long smart card and vehicle location data. The results indicate that the average rider follows the compensatory rule on only 25.5\% of route choice occasions. DLCM estimates also show an increase of 47\% in metro riders' valuation of travel time under extremely crowded conditions relative to that under uncrowded conditions. \\

\noindent \textbf{Keywords:} Dynamic preferences;  Crowding valuation; Inertia; Habit; Expectation-maximization. 

\newpage

\section{Introduction}

\subsection{Empirical Context}
Quantitative measurement of the user valuation of key attributes of the public transport trip, including in-vehicle travel time and crowding, is important in investment appraisal, demand modelling, and supply-side decisions such as fare optimisation. Traditionally, most of the research articles and consultancy reports use a stated preference (SP) survey and estimate the traveller's perceived value of crowding in terms of a \textit{crowding multiplier} -- the ratio of value-of-travel-time under crowded and uncrowded conditions \citep{wardman2011twenty}. SP studies generally elicit preferences of riders in a hypothetical route choice experiment and estimate discrete choice models (DCMs) to obtain the crowding multiplier \citep[See][for the review]{bansal2019flexible}. 

Whereas the hypothetical bias is a major limitation of the SP data, the required information to estimate DCMs (riders' route preferences and attributes of all available routes) is difficult to obtain using conventional revealed preference (RP) surveys \citep{tirachini2016valuation}. Due to these challenges, early crowding valuation studies relying on the RP data either deviated from DCMs \citep{kroes2014value} or complemented the RP data with SP data \citep{batarce2015use}. However, the emerging use of smart cards for fare collection provides an alternative way to collect the required RP data. \cite{tirachini2016valuation} first illustrate how smart card data can be used to estimate the standing penalty of Mass Rapid Transit users in Singapore, i.e.\,the disutility of standing measured in the equivalent travel time loss. \cite{horcher2017crowding} integrate smart card data with automated vehicle location (AVL) data to estimate the crowding multiplier of Hong Kong Mass Transit Railway (MTR) users. 

\subsection{Empirical Research Gaps}  \label{sec:erg}
We identify two research gaps in the crowding valuation literature. First, whereas dynamic route preferences and learning behaviours are hard to capture in the SP experiments, previous RP studies also rely on static choice models. This implies that strong assumptions had to be made on how users form expectations about attribute levels on alternative routes. For example, \citet{horcher2017crowding} assume that the observed travellers are experienced enough to know the \textit{average} train occupancy on available routes, at the time of day when their journeys begin. Second, even if expectations in the choice situation are correctly recovered, a regular subway user, e.g. a daily commuter, might not actively make a compensatory route choice before every trip. In fact, a rider can adhere to the same route until a bad experience occurs, but none of the previous public transport studies model such non-compensatory behaviour. The riders who choose routes based on inertia or habit should not contribute toward the crowding cost valuation and therefore, modelling such behaviour is crucial to accurately estimate the crowding multiplier. 

A dynamic choice model that can identify the temporal variation in the decision rules (compensatory versus non-compensatory) used by a rider can address both research gaps. To explore if any such model exists, we succinctly review the relevant literature on modelling of different decision rules and dynamic choice models in sections \ref{sec:review_dr} and \ref{sec:review_dcm}, respectively.

\subsection{Literature Review: Decision Rules} \label{sec:review_dr}
Most of the previous studies modify the systematic part of the indirect utility to incorporate different decision rules  \citep{elrod2004new, swait2001non}. A few studies have used the latent class specification to simultaneously model multiple decision rules considering one-to-one correspondence between a decision rule and a latent class \citep{dey2018accommodating,hess2012allowing,swait2001influence}. However, these models are static, i.e. they assume that the decision-maker uses the same decision rule across all choice occasions. Moreover, these studies test their latent class models using rather simplistic stated preference datasets.    

\subsection{Literature Review: Dynamic choice Models} \label{sec:review_dcm}
In travel behaviour modelling, cumulative prospect theory (CPT), instance-based learning theory (IBLT), and hidden Markov models (HMMs) are popular approaches to understand the dynamics of travel preferences.   

CPT is particularly used in eliciting day-to-day travel mode \citep{yang2017experimental,ghader2019modeling} and route choices \citep{jou2013application, yang2014development}. These choices fit in the framework of \textit{decision-making under uncertainty} because travellers do not have perfect information of routes' or modes' characteristics, and thus are prone to violate the rationality assumption of the expected utility-maximization theory. Despite the success of CPT in understanding travellers' dynamic preferences, there are two practical concerns in using it. First, the results are highly sensitive to the reference point \citep{avineri2006effect} and the parameters of gain and loss functions \citep{jou2013application}. Ideally, a stated preference survey is required to estimate these parameters. Second, travel choices are experience-based, but CPT is applicable to the description-based decision-making \citep{erev2010choice, jou2013application}.

Unlike CPT, IBLT is appropriate for experience-based learning situations. This psychological theory relies on the power law of forgetting \citep{gonzalez2003instance}. \cite{tang2017exploratory} first illustrated the application of IBLT in dynamic route choice models by integrating it into the econometric framework of the mixed logit model. The proposed IBLT-based model also accounts for the hot stove effect (i.e., bad outcomes have a lasting effect) and the pay-off variability effect (i.e., a larger variability in payoffs leads to random choices).   

HMMs were originally developed in the machine learning literature, but choice modellers also find them appealing because they offer a flexible econometric framework to model dynamic choices. However, similar to IBLT, only a handful of studies have adopted HMMs to estimate and forecast dynamic choices of car ownership \citep{yang2017hidden, xiong2018high} and travel modes \citep{xiong2015analysis, xiong2017dynamic, xiong2018measuring, zarwi2017modeling}. We identify four main limitations of these studies. First, previous studies relying on HMMs do not account for the non-compensatory choice process (e.g., inertia or habit). \cite{cantillo2007modeling}, \cite{cherchi2011accounting}, and \cite{gonzalez2017testing} model inertia and habit dynamics of decision-makers by including the lagged utilities in traditional choice models, but those studies miss the benefits of HMMs. Second, the state-specific distribution ignores the learning of travellers from previous trips, i.e. state-specific distribution is specified using attributes of the current period. Third, previous HMM-based studies use datasets with ten or fewer periods, which is not sufficient to exploit the utility of such dynamic models. This is perhaps the main reason that previous studies do not fully leverage the ability of HMMs to associate a behavioural interpretation to the hidden states (or latent classes). Fourth, the state-specific choice model in existing HMM studies ignores cross-consumer heterogeneity. This constrained specification could cause confusion between heterogeneity and dynamics because some states may capture heterogeneity along with dynamics \citep{netzer2017hidden}. Due to all these challenges, we cannot use off-the-shelf HMMs to formulate the dynamic choice process of subway riders. 

We also note that, except \cite{zarwi2017modeling}, all aforementioned studies estimate HMMs using the Markov Chain Monte Carlo (MCMC) simulation, but none of them discusses the prevalent issue of label switching and subsequent remedies \citep{spezia2009reversible}.

\subsection{Contributions}
Our review suggests that, ironically, the existing choice models accounting for preference dynamics assume a fully-compensatory decision rule and the models incorporating heterogeneity in decision rules are not dynamic. 

In this study, we propose a dynamic latent class model (DLCM) which incorporates the learning behaviour of riders using the IBLT, specifies compensatory and non-compensatory (i.e., inertia/habit) choice processes of riders as latent classes, and allows them to dynamically transition between these classes based on the differences between the expected and the experienced level of services and other historical attributes. Our model can also account for the unobserved heterogeneity in preferences of riders. Thus, the proposed DLCM provides a comprehensive and general framework to model dynamic choices while accounting for heterogeneity in decision rules. The resulting model turns out to be a new variant of the heterogeneous HMM where a rider's choice at any instance not only depends on the rider's current class (i.e., state), but is also influenced by the rider's lagged choice. To circumvent the label switching issue of MCMC, we extend the expectation-maximization (EM) algorithm for HMMs to estimate the proposed DLCM. We also adapt the Viterbi algorithm to predict the most likely sequence of latent classes of a rider, conditional on her observed route choices \citep{arulampalam2002tutorial}.   

We illustrate the applicability of the proposed DLCM in addressing the empirical research gaps (as discussed in section \ref{sec:erg}) by estimating the crowding cost of an Asian metro's riders. In doing so, we calibrate DLCM using a two-month-long dynamic panel dataset on riders' revealed route preferences. This is the first such application in the crowding valuation literature. 

The remainder of this paper is organized as follows: Section \ref{sec:Model} formulates DLCM; Section \ref{sec:est} derives the EM algorithm to estimate DLCM and provides inference procedure; Section \ref{sec:mc} describes the simulation setup and discusses results of the Monte Carlo study to validate the model formulation and estimation. Sections \ref{sec:Exp}-\ref{sec:emp} present an empirical application of DCLM: Section \ref{sec:Exp} discusses the longitudinal data from which route choices and trip experiences of riders are derived, together with the corresponding RP experiment design; Section \ref{sec:emp} illustrates the importance of DLCM by investigating the results of the empirical study. Conclusions and future work are discussed in Section \ref{sec:conc}.          

\section{Model Formulation} \label{sec:Model}

Assume that the researcher records observations of the route choice and trip experience of regular commuters, and these observations can be linked to each other through unique passenger (smart card) identifiers, thus documenting a sequence of repeated choices. The proposed DLCM has three components -- initialisation model, transition model, and choice model. The long panel data allows us to utilize the first few observations of riders to identify their initial latent classes. We consider that a rider can choose to be in any of two latent classes (or hidden states) at a choice occasion: 1) compensatory, 2) non-compensatory (i.e. inclined to make choices due to habit or inertia). In the choice model, conditional on the latent class and the lagged choice, a rider chooses a route from a set of two available routes. In what follows, we formulate DLCM for two alternatives and two latent classes, but without loss of generality, it can be extended to any number of alternatives and latent classes. For simplicity, we first describe transition and initialisation models, followed by the choice model. 

\subsection{Transition Model} 
A rider's class transition probabilities are likely to depend on the difference between the expected and experienced level of service on the route chosen in the previous period. Moreover, a choice sequence of a rider also provides information about the class of the rider. For example, a consistent route choice across several occasions indicate a rider's inclination towards being in the non-compensatory class.

If rider $i$ is in class $s$ at time $t$, the utility $M_{its}$ derived by her due to a mismatch between the expected and the experienced level of service at the chosen route $j_t$ is: 

\begin{equation}
M_{its} = m_{its}  + \epsilon_{its}  = \bm{\zeta}_{s} \left[\bm{X}_{itj_{t}} - \mathbb{E}(\bm{X}_{itj_{t}}) \right]  + \bm{\zeta}_{s}^{C} \bm{X}_{it}^{C} +  \epsilon_{its}, 
\end{equation}

where $\bm{X}_{itj_{t}}$ is a vector of attributes (e.g., crowding level) experienced by rider $i$ on chosen route $j_{t}$ at time $t$. $\bm{X}_{it}^{C}$ is derived from a sequence of choices made by rider $i$. A proportion of choice transitions made by rider $i$ in the choice sequence observed until time $t$ is one such attribute. We define expected values of $\bm{X}_{itj_{t}}$:   $\mathbb{E}(\bm{X}_{itj_{t}})$ using IBLT \citep{tang2017exploratory}:
\begin{equation}\label{eq:expect}
\begin{split}
 & \mathbb{E}(\bm{X}_{itj_{t}})  = \sum_{t^{\prime}} W_{t^{\prime}t} \bm{X}_{it^{\prime}j_{t}}, \quad t^{\prime},\tau \in \{t-1, t-2,...\},\\
 \text{where} \quad & W_{t^{\prime}t}  = \frac{[t-t^{\prime}]^{-\mu}}{{\sum_{\forall \tau}[t-\tau]^{-\mu}}},\\
\end{split}
\end{equation}

where $\mu$ is a memory decay parameter that captures the rate of forgetting the past experiences. Assuming Gumbel distributed $\epsilon_{its}$, transition probability expressions are: 
\begin{equation}
\begin{split}
    P(s_{i(t+1)} = 1 \lvert s_{it}=s;\bm{\zeta}_{s}, \bm{\zeta}_{s}^{C},\mu) &= \frac{\exp\left(m_{its}\right)}{1+ \exp\left(m_{its}\right)}, \\
    P(s_{i(t+1)} = 2 \lvert s_{it}=s; \bm{\zeta}_{s}, \bm{\zeta}_{s}^{C}, \mu) &= 1-P(s_{i(t+1)} = 1 \lvert s_{it}=s;\bm{\zeta}_{s}, \bm{\zeta}_{s}^{C}, \mu),
\end{split}
\end{equation}

If  $\bm{X}_{itj_{t}}$ includes the level-of-service attributes for which ``less is better" (e.g., travel time, crowding), we expect $\bm{\zeta}_{s}$ to be positive. Intuitively, if the experienced level of service is poorer than the prior expectation at time $t$, a rider is more likely to remain in or switch to the compensatory class (class 1) at $t+1$. 

\subsection{Initialisation Model} 
Consider that we observe a rider for $T_{I}+T$ periods. Since we only observe the attributes of the route chosen by a rider, we select $T_I$ in such a way that the rider at least chooses both routes once by the time $T_I$ (see Section \ref{sec:datapro} for details on how the length of this phase is determined in our empirical application). We do not include the first $T_{I}$ choices of a rider in the choice model because a choice model cannot be estimated in the absence of attributes of both routes. If we shift the time clock by $T_{I}$ periods, the latent class and the choice at $t=T_{I}+1$ correspond to those at $t=1$. We thus consider the latent class probabilities \textit{after} the choice made at time $T_{I}$ (i.e., at time $t=T_{I}+1$) as the \textit{initial} latent class probabilities.       

Similar to the transition model, based on the differences between the experienced and expected level of service on the chosen route at $t=T_{I}$, we can obtain the latent class probabilities of a rider after the choice made at $t=T_{I}$ . 

\begin{equation}\label{eq:ini}
\begin{split}
K_{iT_{I}} = k_{iT_{I}}  + \epsilon_{iT_{I}}  & = \bm{\zeta}_{0} \left[\bm{Z}_{iT_{I}j_{T_{I}}} - \mathbb{E}(\bm{Z}_{iT_{I}j_{T_{I}}}) \right]  + \bm{\zeta}_{0}^{C} \bm{Z}_{iT_{I}}^{C} + \epsilon_{iT_{I}}, \\
P(s_{i(T_{I}+1)} = 1 ;\bm{\zeta}_{0},\bm{\zeta}_{0}^{C}, \mu) &= \frac{\exp\left(k_{iT_{I}}\right)}{1+ \exp\left(k_{iT_{I}}\right)}, \\
P(s_{i(T_{I}+1)} = 2; \bm{\zeta}_{0}, \bm{\zeta}_{0}^{C},\mu) &= 1-P(s_{i(T_{I}+1)} = 1 ;\bm{\zeta}_{0},\bm{\zeta}_{0}^{C}, \mu),
\end{split}
\end{equation}

Note that the first $T_{I}-1$ choices of a rider, if at all, are utilized to compute the expected level of service $\mathbb{E}(\bm{Z}_{iT_{I}j_{T_{I}}})$ for the route chosen by the rider at $t=T_{I}$ and creating other attributes $\bm{Z}_{iT_{I}}^{C}$. The expectation is computed using equation \ref{eq:expect}. 

\subsection{Choice model}
If rider $i$ is in the compensatory class at time $t$ (i.e., $s_{it}=1$), her utility from choosing route $j$ at time $t$ is:

\begin{equation}
    U_{itj} = V_{itj} + \nu_{itj} = \bm{\gamma} \mathbb{E}(\bm{F}_{itj}) + \bm{\chi}_{i} \mathbb{E}(\bm{G}_{itj}) + \nu_{itj} , \quad \text{where} \quad \bm{\chi}_{i} \sim \text{Normal}(\bm{\varrho}, \bm{\Psi}) 
\end{equation}

We consider that the marginal utility associated with attributes $\bm{F}_{itj}$ do not vary across riders, but preference heterogeneity is present for attributes $\bm{G}_{itj}$. $\bm{\chi}_{i}$ is assumed to follow normal distribution, but any parametric or semi-parametric mixing distribution can be specified depending on the context. The expected value of attributes is obtained using equation \ref{eq:expect}. If $y_{it}$ is the route chosen by rider $i$ at time $t$ and $\nu_{itj}$ is Gumbel-distributed idiosyncratic error term, the route choice probabilities of passenger $i$, conditional on being in the compensatory class at the beginning of time $t$, is:
\begin{equation}
\begin{split}
    P(y_{it} = 1 \lvert s_{it}=1; \bm{\gamma}, \bm{\varrho}, \bm{\Psi}) &= \frac{\exp\left(V_{it1} \right)}{\exp\left(V_{it1}\right) + \exp\left(V_{it2} \right)}  \\
   P(y_{it} = 2 \lvert s_{it}=1; \bm{\gamma}, \bm{\varrho}, \bm{\Psi}) &= 1 - P(y_{it} = 1 \lvert s_{it}=1; \bm{\gamma}, \bm{\varrho}, \bm{\Psi})
\end{split}
\end{equation}

If a rider is in the non-compensatory class at time $t$, she is more likely to choose the same route at time $t$ as chosen at $t-1$. Based on this observation, we now define the route choice probabilities if rider $i$ is in non-compensatory class at $t$ (i.e., $s_{it}=2$): 
\begin{equation}
\begin{split}
    P(y_{it} = 1 \lvert s_{it}=2, y_{i(t-1)}; \lambda_{1},\lambda_{2}) &= \frac{\exp\left(\lambda_{1} \mathbb{1}[y_{i(t-1)} = 1] - \lambda_{2}\mathbb{1}[y_{i(t-1)} = 2]\right)}{\exp\left(\lambda_{1} \mathbb{1}[y_{i(t-1)} = 1] - \lambda_{2}\mathbb{1}[y_{i(t-1)} = 2]\right) +1} \\
   P(y_{it} = 2 \lvert s_{it}=2, y_{i(t-1)}; \lambda_{1},\lambda_{2}) &= 1 - P(y_{it} = 1 \lvert s_{it}=2, y_{i(t-1)}; \lambda_{1},\lambda_{2})
\end{split}
\end{equation}

 where $\mathbb{1}[.]$ is an indicator function. We would expect $\lambda_1$ and $\lambda_2$ to be highly positive because the passenger is likely to make the same choice in two consecutive scenarios due to inertia or habit. Some route-specific attributes derived from the historical choice sequence of a rider can also be incorporated in the systematic utility. We choose this specific form of choice probabilities for the non-compensatory class based on the context of the empirical study but there is a flexibility to modify this function in other empirical contexts.

\section{Model Estimation}  \label{sec:est}
\begin{figure}[htbp] 
\centering
\includegraphics[width=1\textwidth]{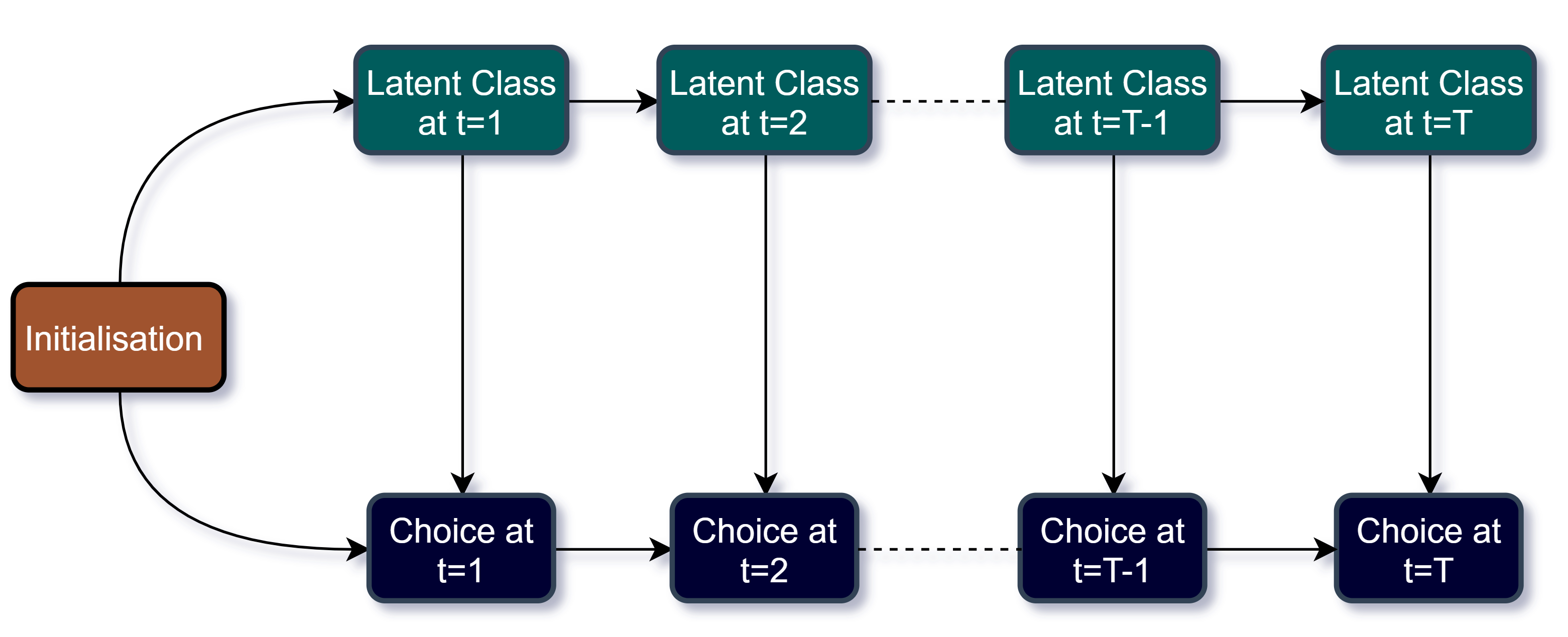}
\caption{The dynamic latent class model} 
\label{fig:sketch}
\end{figure}

By combining all three components of the model, we write the conditional likelihood of the model: 
\begin{equation}
\begin{split}
    P(y_{i1},\dots,y_{iT} \lvert \bm{X}, \bm{Z}, \bm{F}, \bm{G}; \bm{\Theta}) &= \sum_{s_1 = 1}^2\sum_{s_2 = 1}^2 \dots \sum_{s_T = 1}^2 \underbrace{\prod_{t=1}^{T} P\left(y_{it} \lvert q_{its_{t}} = 1, y_{i(t-1)}\right)}_{\text{Choice Model}} \underbrace{P(q_{i1s_{1}} = 1 \lvert \text{Inputs})}_{\text{Initialisation Model}}\dots \\
    &\dots  \underbrace{\prod_{t=2}^{T} P(q_{its_{t}}=1 \lvert q_{i(t-1)s_{(t-1)}}=1)}_{\text{Transition Model}} 
\end{split}
\end{equation}
 
 where $q_{its}$ is 1 if the passenger $i$ belongs to class $s$ at time $t$, else it is zero. The model parameters are $\bm{\Theta} = \left\{\mu, \bm{\zeta}_{0}, \bm{\zeta}_{1},\bm{\zeta}_{2}, \bm{\zeta}_{0}^{C}, \bm{\zeta}_{1}^{C},\bm{\zeta}_{2}^{C}, \bm{\gamma},  \bm{\varrho}, \bm{\Psi}, \lambda_1, \lambda_2 \right\}$. Figure \ref{fig:sketch} shows the schematic diagram of the proposed DLCM. This specification can be viewed as a variant of the traditional heterogeneous hidden Markov models where conditional on the latent class, choice probabilities also depend on the lagged choice.   
 
\subsection{Expectation-Maximization (EM) Algorithm}
Direct maximization of the likelihood is challenging due to a well-known risk of underflow (i.e., the product of probabilities is too small to be represented by the CPU). Moreover, the analytical gradient expression of the likelihood is complex and maximization using numerical gradients can result in prohibitively large computation time \citep{netzer2017hidden}. To decompose the likelihood maximisation into simplified optimisation problems, we estimate DLCM using the EM algorithm. Readers can refer \cite{bansal2018minorization}, \cite{bhat1997endogenous},\cite{sohn2017expectation}, and \cite{zarwi2017modeling} to know more about applications of the EM algorithm in estimating choice models. We extend the existing EM algorithm for the heterogeneous HMMs to account for the auto-correlated choices, preference heterogeneity in the choice model, and riders' learning behaviour.  

The EM algorithm was originally developed to deal with the missing data problem. The DLCM likelihood maximisation problem also falls under the same category because latent classes can be treated as the missing data. The EM algorithm is a two-step iterative procedure where the conditional expectation of the missing data is obtained in the E-step and then the complete loglikelihood is maximised in the M-step to update the model parameters. The convergence criterion is defined based on the difference in parameter estimates or loglikelihood values of two consecutive iterations.  

Assuming the latent class assignment as the missing variable, we write the complete likelihood $L_{c}$ and the complete loglikelihood $\log L_{c}$ of the model:
\begin{equation}
\begin{split}
    L_{c} & = P(\bm{y}_{1},\dots,\bm{y}_{T}, \bm{s}_{1}, \dots, \bm{s}_{T}; \bm{\Theta}) \\
    & = \left[\prod_{i=1}^N \prod_{t=1}^T \prod_{s=1}^2 \left[P\left(y_{it} \lvert q_{its} = 1, y_{i(t-1)}\right)\right]^{q_{its}}\right] \left[ \prod_{i=1}^N \prod_{s=1}^2 \left[P(q_{i1s} = 1 \lvert \text{Inputs})\right]^{q_{i1s}} \right] \dots \\
    & \dots \left[\prod_{i=1}^N \prod_{t=2}^T \prod_{s=1}^2 \prod_{r=1}^2 \left[ P(q_{its}=1 \lvert q_{i(t-1)r}=1) \right]^{q_{its}q_{i(t-1)r}} \right]  \\
    \log L_{c} & = \left[\sum_{i=1}^N \sum_{t=1}^T \sum_{s=1}^2 q_{its} \log \left[P\left(y_{it} \lvert q_{its} = 1, y_{i(t-1)}\right)\right]\right]  \dots \\
    & \dots + \left[ \sum_{i=1}^N \sum_{s=1}^2 q_{i1s} \log \left[P(q_{i1s} = 1 \lvert \text{Inputs})\right]\right] \dots \\
    & \dots + \left[\sum_{i=1}^N \sum_{t=2}^T \sum_{s=1}^2 \sum_{r=1}^2 [q_{its}q_{i(t-1)r}] \log\left[ P(q_{its}=1 \lvert q_{i(t-1)r}=1) \right] \right] 
\end{split}
\end{equation}

\subsubsection{E-step}
Based on the complete loglikelihood $\log L_{c}$ expression, the E-step in $(k+1)^{th}$ iteration requires computing the following expectations:
\begin{equation}
    \begin{split}
        \pi_{its}^{k+1} & = \mathbb{E}[q_{its} \lvert \bm{y}_i;\bm{\Theta}^{k}] = P[q_{its} = 1\lvert \bm{y}_i;\bm{\Theta}^{k}]\\
        \omega_{itrs}^{k+1} & = \mathbb{E}[q_{its}q_{i(t-1)r}\lvert \bm{y}_i;\bm{\Theta}^{k}] = P[q_{its}q_{i(t-1)r} = 1\lvert \bm{y}_i;\bm{\Theta}^{k}]
    \end{split}
\end{equation}

To compute expectations in the E-step efficiently, we define forward $(\alpha_{its})$ and backward $(\beta_{its})$ variables: 
\begin{equation}
    \begin{split}
        \alpha_{its}(\bm{\Theta}) & =  P(y_{i1}, \dots, y_{it}, q_{its} = 1 ; \bm{\Theta}) \\
        \beta_{its}(\bm{\Theta}) & =  P(y_{i(t+1)}, \dots, y_{iT} \lvert y_{it}, q_{its} = 1;  \bm{\Theta})
    \end{split}
\end{equation}

We then compute the $\pi_{its}^{k+1}$ and $\omega_{itrs}^{k+1}$ in terms of forward and backward variables using the Bayes theorem.
\begin{equation}
    \begin{split}
        \pi_{its}^{k+1}  &= P[q_{its}=1 \lvert \bm{y}_i;\bm{\Theta}^{k}] \\
        & = \frac{P(\bm{y}_{i} \lvert  q_{its} =1; \bm{\Theta}^{k}) P(q_{its} =1 ; \bm{\Theta}^{k})}{P(\bm{y}_i ; \bm{\Theta}^{k})}\\
        & = \frac{P(y_{i1},\dots,y_{it},y_{i(t+1)},\dots,y_{iT} \lvert  q_{its} =1; \bm{\Theta}^{k}) P(q_{its} =1; \bm{\Theta}^{k})}{P(\bm{y}_i; \bm{\Theta}^{k})} \\
        & = \frac{P(y_{i1},\dots,y_{it} \lvert  q_{its} =1; \bm{\Theta}^{k}) P(y_{i(t+1)},\dots,y_{iT} \lvert y_{it}, q_{its} =1; \bm{\Theta}^{k}) P(q_{its} =1; \bm{\Theta}^{k})}{P(\bm{y}_i; \bm{\Theta}^{k})} \\
        & = \frac{P(y_{i1},\dots,y_{it}, q_{its} =1; \bm{\Theta}^{k}) P(y_{i(t+1)},\dots,y_{iT} \lvert y_{it}, q_{its} =1; \bm{\Theta}^{k})}{P(\bm{y}_i; \bm{\Theta}^{k})}\\
        &=\frac{\alpha_{its}(\bm{\Theta}^{k}) \beta_{its}(\bm{\Theta}^{k})}{P(\bm{y}_i; \bm{\Theta}^{k})} \\
        &= \frac{\alpha_{its}(\bm{\Theta}^{k}) \beta_{its}(\bm{\Theta}^{k})}{\sum_{s^{\prime}=1}^{2}\alpha_{its^{\prime}}(\bm{\Theta}^{k}) \beta_{its^{\prime}}(\bm{\Theta}^{k})}
    \end{split}
\end{equation}
From now onward, we omit $\bm{\Theta}_{k}$ for brevity. 
\begin{equation}
    \begin{split}
        \omega_{itrs}^{k+1} & = P[q_{its}q_{i(t-1)r} = 1\lvert \bm{y}_i] \\
        & = \frac{P(\bm{y}_i \lvert q_{its}q_{i(t-1)r} = 1) P(q_{its}q_{i(t-1)r} = 1)}{P(\bm{y}_i)} \\
        & = \frac{\left[\splitdfrac{P(y_{i1},\dots,y_{i(t-1)} \lvert q_{i(t-1)r} = 1) P(y_{it} \lvert y_{i(t-1)}, q_{its}=1)}{P(y_{i(t+1)},\dots ,y_{iT} \lvert y_{it}, q_{its} = 1) P(q_{its} = 1 \lvert q_{i(t-1)r} = 1)P(q_{i(t-1)r} = 1)}\right]}{P(\bm{y}_i)}\\
        & = \frac{\alpha_{i(t-1)r}P(y_{it} \lvert y_{i(t-1)}, q_{its}=1) \beta_{its} P(q_{its} =1 \lvert q_{i(t-1)r} = 1)}{\sum_{s=1}^{2}\alpha_{its} \beta_{its}}
    \end{split}
\end{equation}
The computation details of forward $\alpha_{its}$ and backward $\beta_{its}$ variables are provided in the Appendix \ref{app:E}. 

\subsubsection{M-step}
After computing $\pi_{its}^{k}$ and $\omega_{itrs}^{k}$ in the E-step, the complete loglikelihood is maximised to obtain the parameters for $(k+1)^{\text{th}}$ iteration .   
\begin{equation}
\begin{split}
\bm{\Theta}^{k+1} = \operatorname*{arg\,max}_{\bm{\Theta}} & \Bigg[\sum_{i=1}^N \sum_{t=1}^T \sum_{s=1}^2 \pi_{its}^{k}  \log \left[P\left(y_{it} \lvert q_{its} = 1, y_{i(t-1)}\right)\right] \\
& + \sum_{i=1}^N \sum_{s=1}^2 \pi_{i1s}^{k} \log \left[P(q_{i1s} = 1 \lvert \text{Inputs})\right] \\
&  +  \sum_{i=1}^N \sum_{t=2}^T \sum_{s=1}^2 \sum_{r=1}^2 \omega_{itrs}^{k} \log\left[ P(q_{its}=1 \lvert q_{i(t-1)r}=1) \right] \Bigg]
\end{split}
\end{equation}

We adopt the one-dimensional grid search approach to select the value of the IBLT parameter $\mu$ and all other parameters are estimated using the EM algorithm at a given value $\mu$. There are two challenges in estimating the value of $\mu$ using the EM algorithm. First, since all three components of the model include the learning parameter $\mu$, the entire objective function is required to be optimized at once. This defies the purpose of the EM algorithm to decompose a complex optimisation problem into simpler ones. Second, as the value of $\mu$ changes at each iteration of the algorithm, we also need to iteratively update attributes because the expected value of attributes in each component of DLCM depends on $\mu$. Both challenges make the estimation computationally expensive and numerical issues also prevail. However, in our one-dimension grid search strategy, the expected value of attributes are computed only once at the beginning of the algorithm and parameters in the M-step are updated by solving the simpler optimization problems:

\begin{align}
\{\bm{\gamma},  \bm{\varrho}, \bm{\Psi} \}^{k+1} =& \operatorname*{arg\,max}_{\{\bm{\gamma},  \bm{\varrho}, \bm{\Psi} \}}  \Bigg[\sum_{i=1}^N \sum_{t=1}^T \pi_{it1}^{k}  \log \left[P\left(y_{it} \lvert q_{it1} = 1, y_{i(t-1)}\right)\right] \\
\{\lambda_1, \lambda_2 \}^{k+1}  =& \operatorname*{arg\,max}_{\{\lambda_1, \lambda_2 \}}  \Bigg[\sum_{i=1}^N \sum_{t=1}^T \pi_{it2}^{k}  \log \left[P\left(y_{it} \lvert q_{it2} = 1, y_{i(t-1)}\right)\right] \\
\left\{\bm{\zeta}_{0},\bm{\zeta}_{0}^{C}\right\}^{k+1} = &  \operatorname*{arg\,max}_{\left\{\bm{\zeta}_{0},\bm{\zeta}_{0}^{C}\right\}} \sum_{i=1}^N \sum_{s=1}^2 \pi_{i1s}^{k} \log \left[P(q_{i1s} = 1 \lvert \text{Inputs})\right] 
\end{align}

\begin{align}
\left\{\bm{\zeta}_{1},\bm{\zeta}_{1}^{C}\right\}^{k+1}  = &  \operatorname*{arg\,max}_{\left\{\bm{\zeta}_{1},\bm{\zeta}_{1}^{C}\right\}} \sum_{i=1}^N \sum_{t=2}^T \sum_{s=1}^2  \omega_{it1s}^{k} \log\left[ P(q_{its}=1 \lvert q_{i(t-1)1}=1) \right]  \\
\left\{\bm{\zeta}_{2},\bm{\zeta}_{2}^{C}\right\}^{k+1} = &  \operatorname*{arg\,max}_{\left\{\bm{\zeta}_{2},\bm{\zeta}_{2}^{C}\right\}} \sum_{i=1}^N \sum_{t=2}^T \sum_{s=1}^2  \omega_{it2s}^{k} \log\left[ P(q_{its}=1 \lvert q_{i(t-1)2}=1) \right] 
\end{align}

Whereas the update for $\{\bm{\gamma},  \bm{\varrho}, \bm{\Psi}\}$ is equivalent to estimating a weighted mixed multinomial logit model, other updates are analogous to the estimation of weighted multinomial logit models. Even after this simplification, the estimation is computationally expensive. For instance, if the EM takes $1000$ iterations to converge, the estimation of DLCM involves the estimation of $1000$ mixed logit models. For this reason, we rely
on the Broyden–Fletcher–Goldfarb–Shanno (BFGS) algorithm with analytical gradients to optimise these functions in Python \citep[see][for analytical gradients of the loglikelihood of the mixed logit model]{bansal2019flexible}.  

\subsubsection{Standard Errors}
In the EM estimation of discrete choice models, the information matrix is generally obtained by taking cross-product of  M-step score vectors at convergence \citep{train2008algorithms}. However, \cite{bansal2018minorization} illustrate that standard errors estimated using this procedure can be biased. Therefore, we only use the EM algorithm to get the point estimates of parameters and obtain standard errors by numerically computing the Hessian of true conditional likelihood at these parameter estimates. We use \textit{numdifftools} library in python to numerically compute the Hessian.      

\subsection{Sequence of Latent Classes}
Previous transport studies relying on HMMs did not focus on estimating the most probable sequence of latent class (i.e., hidden states) for a decision-maker, perhaps due to a lack of behavioural interpretation of classes and short panel datasets. However, the estimation of latent classes is meaningful in this study because a rider's choice behaviour (compensatory vs. non-compensatory) is characterised by these classes. This is even more important for supply-side transit policies because crowding multiplier identified from compensatory choices can be scaled down with the knowledge about the extent of a rider's non-compensatory behaviour.

Conditional on the sequence of observed route choices and parameter estimates, we estimate the most likely sequence of a rider's latent classes. To this end, we use adapt the Viterbi algorithm, which uses forward-backward recursion \citep{arulampalam2002tutorial, forney1973viterbi, he1988extended}. Once we condition on the lagged choices in the recursion, the Viterbi algorithm for the heterogeneous HMMs can be used for the proposed DLCM.   

\section{Monte Carlo Study} \label{sec:mc}
To assess the recovery of parameters and the potential convergence issues in the EM estimation, we present a Monte Carlo study. Since standard errors are calculated using the regular asymptotic theory, we are certain that standard deviation of the parameter estimates across resamples would be close to their asymptotic standard errors and therefore, one sample of the data generating process (DGP) is sufficient for our purposes. We consider a DGP with preference heterogeneity in compensatory class.

\begin{table}[h!]
		\centering
		\caption{Results of the Monte Carlo Study with Unobserved Heterogeneity}
		\label{tab:MCR}
		\begin{adjustwidth}{0cm}{}
		\resizebox{1\textwidth}{!}{
\begin{tabular}{lccccc}
\hline
 & \textbf{True value} & \textbf{Estimated value} & \textbf{Std. err.} & \textbf{z-value} & \textbf{\begin{tabular}[c]{@{}c@{}}Gradient at \\ convergence\end{tabular}} \\ \hline
\multicolumn{6}{l}{\textbf{Initialisation}} \\ \hline
$\zeta_{0}^{1}$ & -1 & -0.98 & 0.17 & -5.7 & 2.14E-03 \\
$\zeta_{0}^{2}$ & 1.1 & 1.16 & 0.41 & 2.8 & -1.43E-04 \\
$\zeta_{0}^{3}$ & 0.9 & 1.12 & 0.41 & 2.8 & 6.50E-05 \\ \hline
\multicolumn{6}{l}{\textbf{Transition Model (compensatory class)}} \\ \hline
$\zeta_{1}^{1}$ & -1 & -1.11 & 0.26 & -4.3 & 2.61E-02 \\
$\zeta_{1}^{2}$ & 1.4 & 1.13 & 0.59 & 1.9 & -1.72E-03 \\
$\zeta_{1}^{3}$ & 1.5 & 1.16 & 0.64 & 1.8 & -5.50E-03 \\ \hline
\multicolumn{6}{l}{\textbf{Transition Model (non-compensatory class)}} \\ \hline
$\zeta_{2}^{1}$ & -1.5 & -1.49 & 0.13 & -11.2 & 1.37E-02 \\
$\zeta_{2}^{2}$ & 1.2 & 1.06 & 0.18 & 5.8 & 7.58E-04 \\
$\zeta_{2}^{3}$ & 1.1 & 0.99 & 0.18 & 5.5 & 3.70E-03 \\ \hline
\multicolumn{6}{l}{\textbf{Choice Model (compensatory class)}} \\ \hline
$\gamma^{1}$ & -1 & -1.16 & 0.24 & -4.8 & 3.07E-04 \\
$\gamma^{2}$ & -1.5 & -1.41 & 0.27 & -5.3 & 5.97E-04 \\
$\varrho^{1}$ & 1.5 & 1.71 & 0.24 & 7.1 & -1.82E-03 \\
$\varrho^{2}$ & -1.5 & -1.67 & 0.24 & -7.1 & 2.08E-03 \\
$\Psi^{11}$ & 1 & 1.22 & 0.58 & 2.1 & -8.21E-04 \\
$\Psi^{22}$ & 1 & 0.93 & 0.63 & 1.5 & 6.98E-04 \\ \hline
\multicolumn{6}{l}{\textbf{Choice Model (non-compensatory class)}} \\ \hline
$\lambda_{1}$ & 1 & 1.02 & 0.03 & 29.8 & 3.68E-03 \\
$\lambda_{2}$ & 2 & 1.98 & 0.08 & 25.0 & 8.03E-03 \\ \hline
Number of observations & \multicolumn{5}{c}{2000} \\
Choice occasions for initialisation & \multicolumn{5}{c}{10} \\
Available choice occasions & \multicolumn{5}{c}{20} \\
Number of EM iterations & \multicolumn{5}{c}{719} \\
Loglikelihood at convergence & \multicolumn{5}{c}{-21513.9} \\
True Loglikelihood & \multicolumn{5}{c}{-21519.9} \\ \hline
\end{tabular}}
\end{adjustwidth}
\end{table}

We generate  each component of explanatory variables $\{\bm{X}, \bm{Z}, \bm{F}, \bm{G}\}$ by taking draws from a normally-distributed random variable with mean $1.5$ and standard deviation of $0.3$. We utilize the first ten choices (i.e., $T_{I} = 10$) of riders to compute initial latent class probabilities and assume that riders develop an expectation for the level of service on a route based on their past three trips on that route. In both DGPs, we consider the memory decay parameter of the IBLT $\mu$ to be 1. We consider $2000$ riders (i.e., $N=2000$) and $30$ choice occasions per rider (i.e., $T+T_I =30$). A diagonal variance-covariance matrix is assumed on random parameters. The algorithm terminates when the absolute difference between the loglikelihood values of two consecutive iterations is below $10^{-6}$. We also try tighter convergence criteria, but results remain consistent.     

Table \ref{tab:MCR} presents the estimation results where superscripts on variables relate to the component number of the vector or matrix. For example, $\zeta_{0}^{2}$ denotes the second element of the vector $\bm{\zeta}_{0}$ and $\Psi^{22}$ implies variance of the second random parameter. A comparison of true and estimated values of parameters indicate that all model parameters are recovered well. Similar values of the loglikelihood at convergence and true loglikelihood further validate the estimation procedure. Gradient values at convergence are also close to zero for all parameters, which ensures the convergence of the EM to a local optimal and discard possibilities of any numerical or identification issues. Since true latent classes (or hidden states) of riders are known in the DGP, we could analyse the performance of the Viterbi algorithm in predicting latent classes. The results indicate that the Viterbi algorithm could predict latent classes correctly at 80.53\% accuracy. The estimation code for this simulation study is provided as supplementary material. 

\section{Data and Experiment Design} \label{sec:Exp}

\begin{figure}[H]
  \centering
  \includegraphics[width=0.8\textwidth]{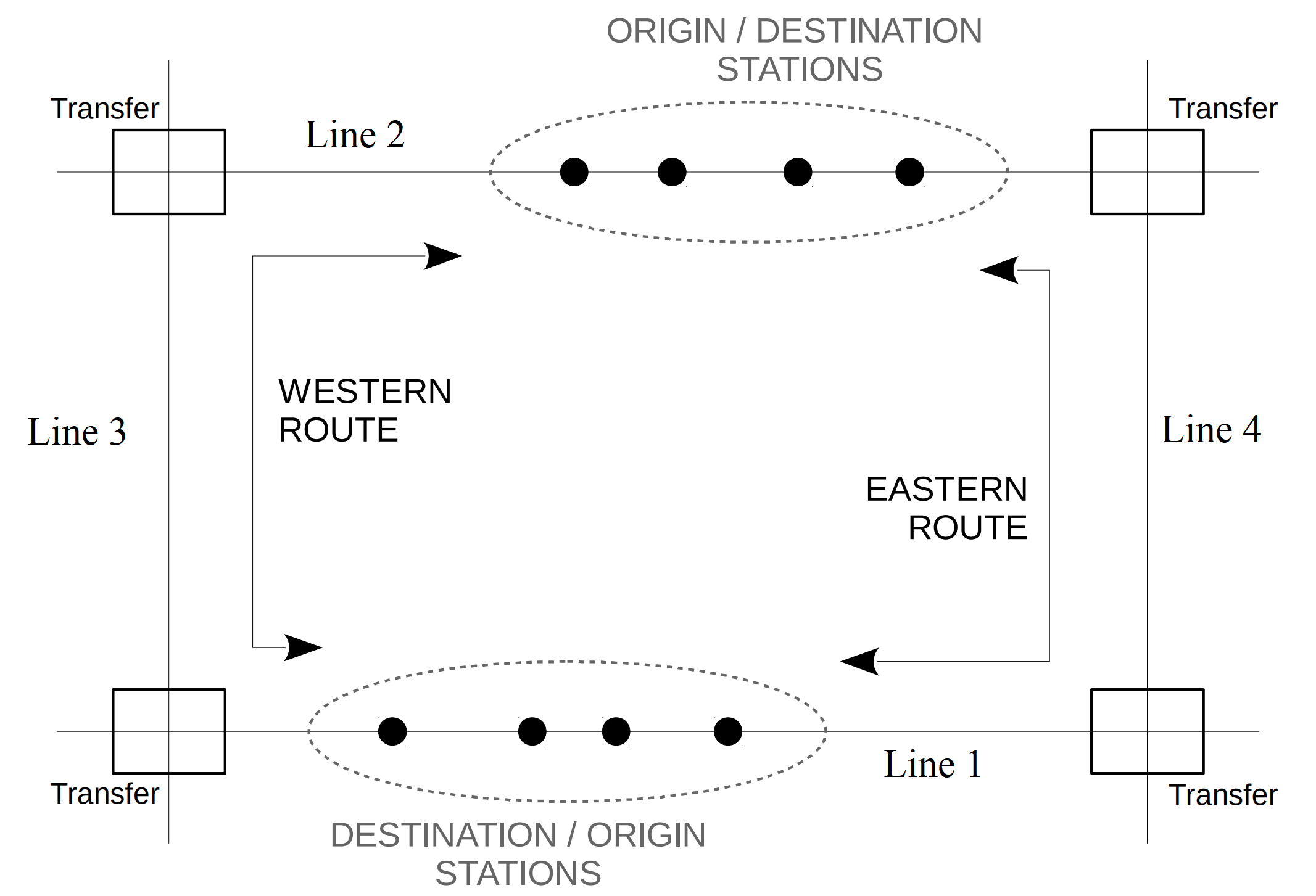}
  \caption{Schematic network layout of the metro network}
  \label{routemap}
\end{figure}

We implement the proposed DLCM using the data from a RP-based route choice experiment. The experiment exploits a unique feature of the network of an Asian metro. Four urban metro lines form an inner-city loop and thus an excellent laboratory for revealed preference route choice data collection. We select 32 origin-destination (OD) pairs between Line 1 and Line 2. Passengers on the selected OD pairs can reach to destinations using exactly two competitive paths, either by transferring to line 3, or to line 4 (see Figure \ref{routemap} for a visual illustration). These paths have enough relative variation in travel time and crowding, circumventing the concern of the dominant alternative. 

\subsection{Recovering travel experience from automated data}

The routes chosen by passengers and attributes of routes are obtained by passing the day-to-day data on automated fare collection (AFC) and vehicle location (AVL) through a passenger-to-train assignment algorithm. To recover the crowding experience of observed travellers throughout the entire length of their journeys, we run a full network assignment and derive in-vehicle crowding densities by aggregating the number of individual passengers travelling on each train. The assignment algorithm follows the methodology introduced by \citet{horcher2017crowding}. However, for the purpose of this experiment, we realise a series of efficiency improvements in the R implementation of the assignment algorithm, in order to make the assignment feasible for the two-month period. This is achieved by making the recovery of trip-level feasible train itineraries quicker using the \textit{Fast Overlap Joins} function available from version 1.12.8 (released in December 2019) of the \textit{data.table} package of R, and by assigning passengers simply to the most probable itinerary instead of the original stochastic assignment of \citet{horcher2017crowding}. With these amendments, the computation time of a one-day assignment decreases from the original two days to just around 25 minutes. 

Key trip attributes include travel time, the density of standing passengers, and the probability of standing. Our algorithm infers the probability of standing on the level of origin-destination pairs. The algorithm is detailed in Section 3.4.3 of \citet{horcher2017crowding}. Both the density of crowding and the probability of standing are recovered for each inter-station section of an experimental passenger's trip. As fares are not differentiated based on the route chosen, we derive crowding cost valuations in terms of the equivalent travel time loss \citep[see the evolution of crowding-dependent value of time multipliers reviewed by][]{wardman2011twenty}. This implies that the trip attributes in DLCM are interactions between travel time and crowding characteristics, and therefore we aggregate the link-level products of train movement times and crowding density or standing probability estimates. Our dataset covers two months, thus allowing for numerous repeated route choice observations from uniquely identified (but otherwise anonymised) smart card holders.

\subsection{Data processing for DLCM} \label{sec:datapro}

The observation unit in our analysis is a rider who travels between a specific origin-destination pair. The smart card number helps us in keeping track of the rider's route choices at several instances. Similar to other revealed preference datasets, we only observe attributes (e.g., travel time and crowding level) of the chosen route. We consider that a rider uses IBLT to develop the expectation of attributes on a route based on her previous experiences on that route. A modeller can only know the rider's expectation of attributes on both routes, if she has chosen them at least once in the past. In the absence of knowledge about (expectation of) attributes of both routes, estimation of a compensatory choice model is infeasible. Therefore, we define the initialisation period $T_{I}$ for each rider based on the time until both alternatives are chosen by a rider. For example, if we observe a rider at ten occasions with the following route choice vector $\{1,1,2,1,1,2,1,1,2,1\}$, we use four occasions for initialisation $(T_{I}=4)$ and the remaining six occasions for choice and transition models $(T=6)$. We follow four sequential screening steps to obtain an appropriate sample for the analysis. 
\begin{enumerate}
    \item We only include riders with at least five observed choices over two months. This criterion is satisfied by 20,960 riders in the population.
    \item We discard riders who frequently travel between two different origin-destination pairs in the same direction (i.e., on the same route) to avoid the mixed learning and contamination of attribute expectations. Specifically, if a rider travels more than two times on other OD pairs in the same direction of the most travelled OD pair, the rider is excluded from the analysis. We are left with 16,328 riders after this filtering. 
    \item The riders who choose only one route or transition to the least-chosen route only once across observed choice occasions are discarded from the analysis. Not to our surprise, this criterion eliminates 13,305 riders (81.4\%), leaving us with 3,023 riders for further analysis. This observation empirically validates our hypothesis that a large proportion of riders follow non-compensatory decision rules (e.g., inertia/habit) while choosing subway routes.    
    \item To ensure that the estimation of transition model utilises preferences of a rider, the rider should at least have two available occasions $(T\geq2)$ after eliminating initial choice instances $T_{I}$. After applying other minor filtering criteria, we are left with 1,947 riders and 1,921 riders, respectively, when we consider one (memory=1) and two past choice occasions (memory=2) in the computation of attributes using IBLT\footnote{We also conduct analysis assuming contribution of the past three choices (memory=3) on a route in computing IBLT-based expectations, but we do not present its results here because lower memory specifications result in better model fit and thus, could better explain the choices made by riders.} (see equation \ref{eq:expect}).
\end{enumerate}

\begin{figure}[h!] 
\centering
\includegraphics[width=0.7\textwidth]{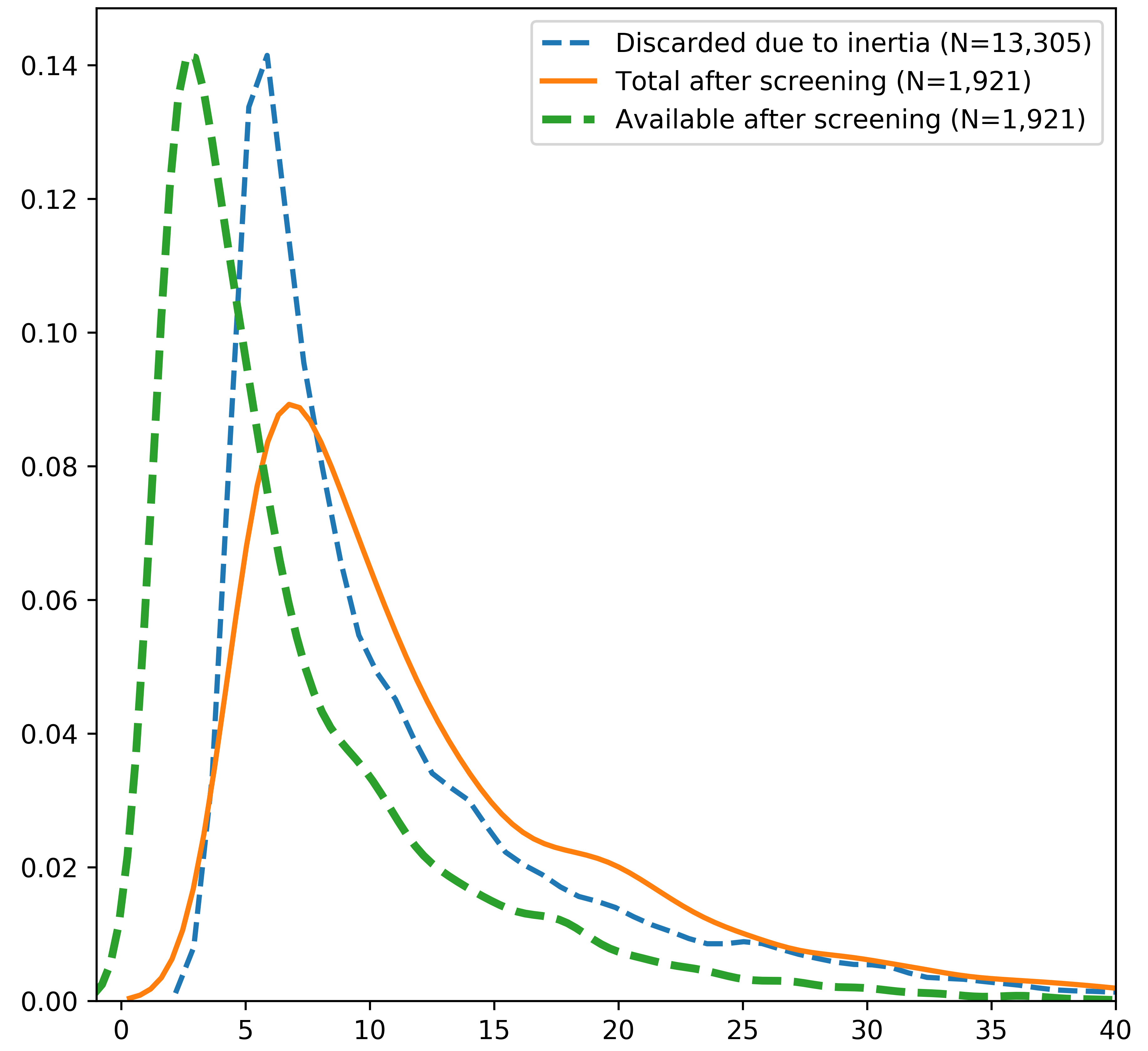}
\caption{Kernel density plot of the number of choice occasions} 
\label{fig:sample}
\end{figure}

To evaluate whether the riders following inertia/habit-based decision rules (i.e., the ones eliminated in step 3) travel more frequently as compared to those in our final sample, we create a kernel density plot of total choice occasions for both in Figure \ref{fig:sample}. Similarities in density plots contradict such hypothesis; in fact, the riders making choices entirely based on habit/inertia appear to be less frequent. In the same figure, we also add a kernel density plot of the \textit{available} choice occasions in the final sample. The added density plot is shifted to the left due to the removal of \textit{initial} choice situations.     

We consider crowding density (measured in passengers per m$^{2}$) and standing probability as dummy (or indicator) covariates in all models. We discretize the parameter space of these covariates because their marginal utilities, when entered in continuous form, do not exhibit statistical significance. Since empirical cumulative density function (CDF) of crowding density has values of 0.39, 0.76, and 0.97 at crowding density of 1, 2, and 3 riders/m$^{2}$,respectively, we create two indicators --  crowding density between 1 and 2 riders/me$^{2}$ (crowding1) and above 2 riders/m$^{2}$ (crowding2), with the base crowding density below 1 rider/m$^{2}$. Similarly, since empirical CDF of standing probability has values of 0.245 and 0.687 at standing probability of 0.4 and 0.7, we create two indicators -- standing probability between 0.4 and 0.7 (SP1) and above 0.7 (SP2), with the base standing probability below 0.4. 

\section{Empirical Results} \label{sec:emp}

\subsection{Results of Static Models} \label{sec:results} 
To highlight the challenges in modelling a dynamic choice process using static models, we estimate them and present their results.

We first estimate the multinomial logit (MNL) model using the data created to estimate the choice component of DLCM. The MNL estimates for memories 1 and 2 are presented in Table \ref{tab:MNL}. While computing expectation of route-specific attributes using IBLT, the memory decay parameter $\mu$ is considered to be 1. For both samples, the coefficient of travel time has an intuitive sign, but the sign of its interaction with SP2 is counter-intuitive -- positive with very high z-value. The results remain virtually the same for different values of $\mu$. The sample also includes riders who follow non-compensatory (inertia-based) decision rules and modelling their choices using a compensatory model might have resulted in this discrepancy.  

\begin{table}[h!]
		\centering
		\caption{Results of the Multinomial Logit Model (MNL)}
		\label{tab:MNL}
		\begin{adjustwidth}{0cm}{}
		\resizebox{1\textwidth}{!}{
		 \begin{threeparttable}
\begin{tabular}{lcccc} \hline
 & \multicolumn{2}{c}{\textbf{Memory = Last choice occasion}} & \multicolumn{2}{c}{\textbf{Memory = Last two choice occasions}} \\ \hline
 & \textbf{Estimate} & \textbf{z-value} & \textbf{Estimate} & \textbf{z-value} \\ \hline
Travel Time (in hours) & -23.30 & -54.5 & -22.69 & -52.1 \\
Travel Time $\times$ crowding1\tnote{a}  $\; \times \;$  SP1\tnote{c} & -0.41 & -3.5 & -0.36 & -3.0 \\
Travel Time $\times$ crowding1  $\times$  SP2 & 0.28 & 1.5 & 0.62 & 3.3 \\
Travel Time $\times$ crowding2  $\times$  SP1 & -0.36 & -1.5 & -0.26 & -1.1 \\
Travel Time $\times$ crowding2\tnote{b}  $\; \times \;$  SP2\tnote{d} & 0.23 & 1.4 & 0.35 & 2.1 \\
Constant (route 1) & -0.19 & -6.6 & -0.17 & -6.0 \\ \hline
Number of observations & \multicolumn{2}{c}{1947} & \multicolumn{2}{c}{1921} \\
Total available choice occasions\tnote{e} & \multicolumn{2}{c}{13447} & \multicolumn{2}{c}{13282} \\
Loglikelihood at convergence & \multicolumn{2}{c}{-6247.5} & \multicolumn{2}{c}{-6303.8} \\\hline
\end{tabular}	
		\begin{tablenotes}
            \item[a] \textbf{crowding1:} dummy variable for the crowding level between 1 and 2 riders/metre$^2$. 
            \item[b] \textbf{crowding2:} dummy variable for the crowding level above 2 riders/metre$^2$. 
			\item[c] \textbf{SP1:} dummy variable for the standing probability between 0.4 and 0.7.                
            \item[d] \textbf{SP2:} dummy variable for the standing probability above 0.7.                        
            \item[e] \textbf{Total available choice occasions:} total choice occasions used in the choice model (sum of $T$ for all riders).
        \end{tablenotes}
        \end{threeparttable}
}
\end{adjustwidth}
\end{table}

With the possibility to segment riders following different decision rules, we estimate the latent class MNL (LC-MNL) model and results are presented in Table \ref{tab:LCMNL}. The improved likelihood suggests that LC-MNL explains the choice process slightly better, but both sign and magnitude of marginal utilities of interaction terms remain counter-intuitive. 

\begin{table}[h!]
		\centering
		\caption{Results of the Latent Class Multinomial Logit Model (LC-MNL)}
		\label{tab:LCMNL}
		\begin{adjustwidth}{0cm}{}
		\resizebox{1\textwidth}{!}{
		 \begin{threeparttable}
\begin{tabular}{lcccccccc} \hline
 & \multicolumn{4}{c}{\textbf{Memory = Last choice occasion}} & \multicolumn{4}{c}{\textbf{Memory = Last two choice occasions}} \\ \hline
 & \multicolumn{2}{c}{\textbf{Class 1}} & \multicolumn{2}{c}{\textbf{Class 2}} & \multicolumn{2}{c}{\textbf{Class 1}} & \multicolumn{2}{c}{\textbf{Class 2}} \\  \hline
 & \textbf{Estimate} & \textbf{z-value} & \textbf{Estimate} & \textbf{z-value} & \textbf{Estimate} & \textbf{z-value} & \textbf{Estimate} & \textbf{z-value} \\  \hline
Travel Time (in hours) & -28.99 & -42.1 & -18.85 & -25.3 & -27.93 & -38.7 & -19.09 & -25.7 \\
Travel Time $\times$ crowding1\tnote{a}   $\; \times \;$  SP1\tnote{c}  & -0.65 & -2.7 & -0.32 & -1.5 & -0.78 & -3.1 & -0.28 & -1.3 \\
Travel Time $\times$ crowding1  $\times$  SP2 & -0.24 & -0.7 & -0.02 & 0.0 & 0.09 & 0.2 & 0.08 & 0.2 \\
Travel Time $\times$ crowding2  $\times$  SP1 & -0.13 & -0.3 & -0.75 & -1.8 & -0.18 & -0.4 & -0.32 & -0.8 \\
Travel Time $\times$ crowding2\tnote{b}   $\; \times \;$  SP2\tnote{d}  & 0.32 & 1.0 & -0.26 & -0.9 & 0.85 & 2.3 & -0.42 & -1.3 \\
Constant (route 1) & 0.92 & 17.2 & -1.51 & -22.5 & 1.01 & 18.1 & -1.47 & -22.7 \\
Class-probability parameter & -0.53 & -24.2 &  &  & -0.48 & -22.4 &  &  \\  \hline
Number of observations & \multicolumn{4}{c}{1947} & \multicolumn{4}{c}{1921} \\
Total available choice occasions\tnote{e}  & \multicolumn{4}{c}{13447} & \multicolumn{4}{c}{13282} \\
Loglikelihood at convergence & \multicolumn{4}{c}{-5387.9} & \multicolumn{4}{c}{-5375.0} \\  \hline
\end{tabular}	
		\begin{tablenotes}
            \item[a] \textbf{crowding1:} dummy variable for the crowding level between 1 and 2 riders/metre$^2$. 
            \item[b] \textbf{crowding2:} dummy variable for the crowding level above 2 riders/metre$^2$. 
			\item[c] \textbf{SP1:} dummy variable for the standing probability between 0.4 and 0.7.                
            \item[d] \textbf{SP2:} dummy variable for the standing probability above 0.7.
            \item[e] \textbf{Total available choice occasions:} total choice occasions used in the choice model (sum of $T$ for all riders).
        \end{tablenotes}
        \end{threeparttable}
}
\end{adjustwidth}
\end{table}

\subsection{Results of the Dynamic Latent Class Model}
We estimate the Dynamic Latent Class Model (DLCM) for both memory values. In the final specification, we keep covariates with z-value greater than 1 and set the EM convergence criterion to $10^{-8}$. 
Since parameter estimates are not very sensitive to the value of memory decay parameter $\mu$, we first set $\mu$ to 1 and find the model specification. Conditional on this specification, we obtain the optimal value of $\mu$ through grid search. 

The parameter estimates of the proposed DLCM  for $\mu=1$ are shown in Table \ref{tab:DLC}. Gradient values of all parameters at convergence are close to zero in both specifications, ensuring convergence to a local optimal. Whereas both specifications have the same number of parameters, the model with the memory of two previous choice occasions explains choices slightly better, i.e. converged to a better loglikelihood (-3734.9 vs. -3740.6) and that too with the rather lower number of \textit{available} choice occasions ($T=13282$ vs. $T=13447$). Therefore, we find the value of $\mu$ through grid search for DLCM with the memory of two previous choice occasions. 

\begin{figure}[h!] 
\centering
\includegraphics[width=0.7\textwidth]{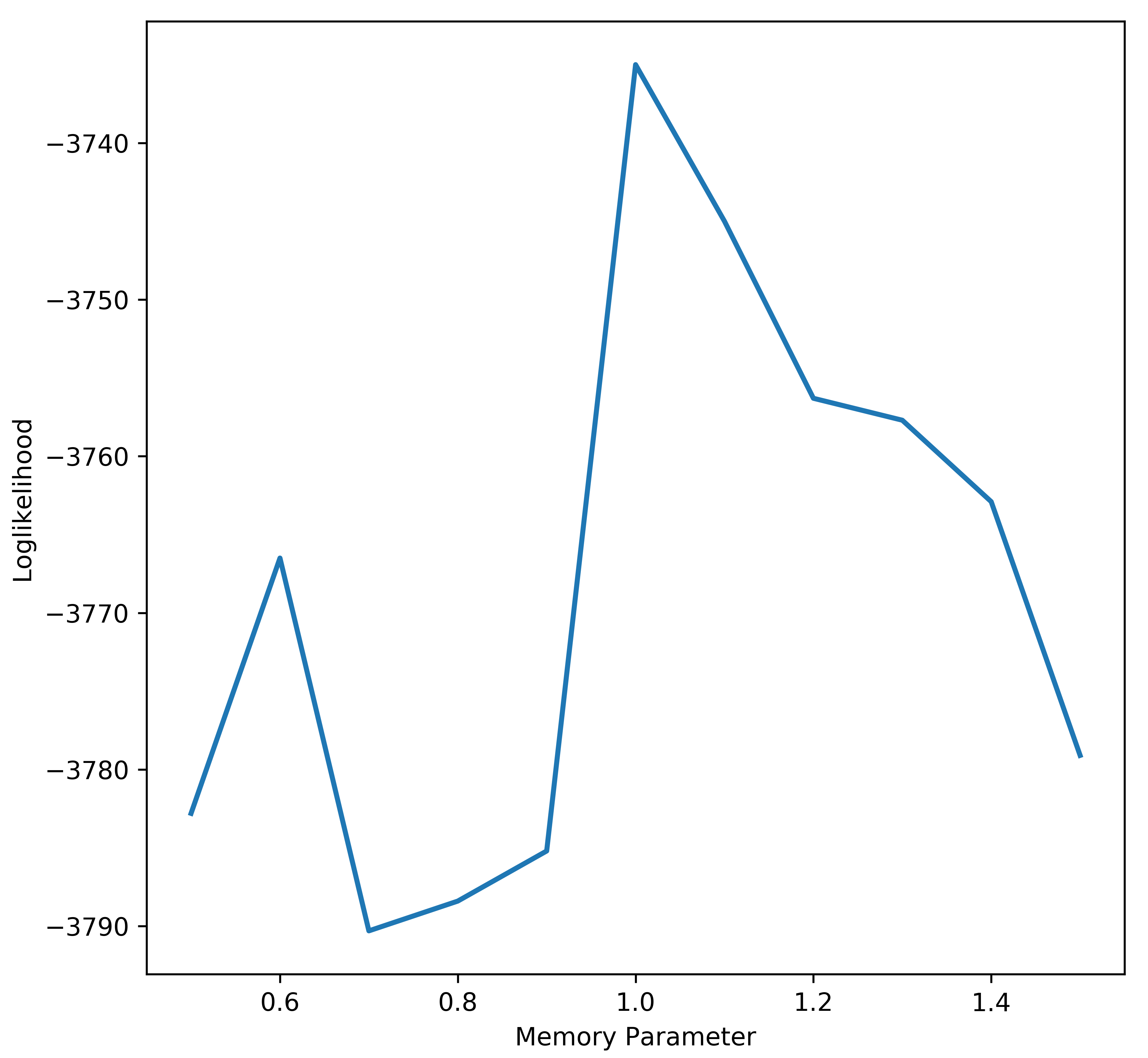}
\caption{One-dimensional grid search to select the memory decay parameter ($\mu$)} 
\label{fig:mu}
\end{figure}

To find the optimal value of $\mu$, we create a one-dimension grid between 0.5 to 1.5, at an increment of 0.1. The resulting loglikelihood values at each grid point are presented in Figure \ref{fig:mu}. The results indicate that the optimal value of $\mu$ is 1 and therefore, the specification presented in Table \ref{tab:DLC} remains the \textit{final} specification.   

\begin{table}[h!]
		\centering
		\caption{Results of the proposed Dynamic Latent Class Model ($\mu=1$)}
		\label{tab:DLC}
		\begin{adjustwidth}{0cm}{}
		\resizebox{1\textwidth}{!}{
		\begin{threeparttable}
\begin{tabular}{lcccccc}
\hline
 & \multicolumn{3}{c}{\textbf{Memory = Last choice occasion}} & \multicolumn{3}{c}{\textbf{Memory = Last two choice occasions}} \\ \hline
 & \textbf{Estimate} & \textbf{z-value} & \textbf{\begin{tabular}[c]{@{}c@{}}Gradient at \\ convergence\end{tabular}} & \textbf{Estimate} & \textbf{z-value} & \textbf{\begin{tabular}[c]{@{}c@{}}Gradient \\ at convergence\end{tabular}} \\ \hline
\textbf{Initialisation} & \multicolumn{1}{l}{} & \multicolumn{1}{l}{} & \multicolumn{1}{l}{} & \multicolumn{1}{l}{} & \multicolumn{1}{l}{} & \multicolumn{1}{l}{} \\ \hline
Constant & 14.55 & 4.7 & 4.00E-06 & 11.66 & 3.7 & 1.80E-05 \\
Highest choice proportion\tnote{a} & -21.27 & -6.2 & -2.00E-06 & -18.15 & -5.4 & 2.00E-06 \\
Proportion of total transitions\tnote{b} & 5.53 & 2.9 & 1.20E-05 & 6.65 & 3.4 & -5.00E-06 \\ \hline
\multicolumn{3}{l}{\textbf{Transition Model (compensatory class)}}  &  &  &  &  \\ \hline
Constant & -2.61 & -4.8 & -5.05E-04 & -3.45 & -4.4 & -8.70E-05 \\
Standing probability & 1.17 & 1.3 & 2.40E-05 & 1.37 & 1.0 & 3.00E-05 \\
Proportion of last transitions\tnote{c} & -14.80 & -4.4 & -4.09E-04 & -27.41 & -4.3 & 1.41E-04 \\
Proportion of total transitions & 24.17 & 6.0 & -2.64E-04 & 38.30 & 5.0 & 4.30E-05 \\ \hline
\multicolumn{3}{l}{\textbf{Transition Model (non-compensatory class)}}  &  &  &  &  \\ \hline
Constant & -5.09 & -13.2 & 9.10E-05 & -5.21 & -14.0 & 4.60E-05 \\
Standing probability & 1.72 & 1.9 & 2.00E-06 & 0.88 & 1.1 & -1.50E-05 \\
Proportion of last transitions & -58.92 & -9.8 & 6.30E-05 & -57.18 & -9.1 & -8.00E-06 \\
Proportion of total transitions & 72.69 & 10.7 & 4.90E-05 & 71.77 & 10.0 & 9.00E-06 \\ \hline
\multicolumn{3}{l}{\textbf{Choice Model (compensatory class)}} &  &  &  &  \\ \hline
Travel Time (in hours) & -5.23 & -8.1 & 3.00E-06 & -5.00 & -7.6 & 9.00E-06 \\
Travel Time $\times$ crowding1\tnote{d}  $\; \times \;$  SP1\tnote{f} & -0.22 & -1.1 & -2.00E-06 & -0.29 & -1.5 & -8.00E-06 \\
Travel Time $\times$ crowding1  $\times$  SP2 & -0.49 & -1.7 & 3.00E-06 & -0.44 & -1.4 & 6.00E-06 \\
Travel Time $\times$ crowding2  $\times$  SP1 & -0.65 & -1.7 & 1.10E-05 & -0.79 & -2.1 & 4.00E-06 \\
Travel Time $\times$ crowding2\tnote{e}  $\; \times \;$  SP2\tnote{g} & -0.87 & -3.3 & 1.70E-05 & -0.97 & -3.4 & -1.20E-05 \\
Constant (route 1) & -0.10 & -2.1 & 4.90E-05 & -0.06 & -1.4 & -4.50E-05 \\ \hline
\multicolumn{3}{l}{\textbf{Choice Model (non-compensatory class)}}  &  &  &  &  \\ \hline
Lagged choice\tnote{h} & 1.49 & 14.2 & -1.44E-04 & 1.50 & 18.3 & 2.70E-05 \\
Choice proportion for a route\tnote{i} & 5.83 & 18.0 & -7.80E-05 & 5.38 & 22.0 & 6.00E-05 \\
Constant (route 1) & -0.13 & -1.4 & -3.10E-05 & -0.11 & -1.3 & 2.00E-05 \\ \hline
Number of observations & \multicolumn{3}{c}{1947} & \multicolumn{3}{c}{1921} \\
Total available choice occasions \tnote{j} & \multicolumn{3}{c}{13447} & \multicolumn{3}{c}{13282} \\
Loglikelihood at convergence & \multicolumn{3}{c}{-3740.6} & \multicolumn{3}{c}{-3734.9} \\ \hline
\end{tabular}
		\begin{tablenotes}
		    \item[a] \textbf{Highest choice proportion:} number of times most chosen route is chosen / $(T_{I} + T)$.
		    \item[b] \textbf{Proportion of total transitions:} total number of route transitions / $(T_{I} + T)$. 
		    \item[c] \textbf{Proportion of last transitions:} number of route transitions until time $t$/ $t$. 
            \item[d] \textbf{crowding1:} dummy variable for the crowding level between 1 and 2 riders/metre$^2$. 
            \item[e] \textbf{crowding2:} dummy variable for the crowding level above 2 riders/metre$^2$. 
			\item[f] \textbf{SP1:} dummy variable for the standing probability between 0.4 and 0.7.                
            \item[g] \textbf{SP2:} dummy variable for the standing probability above 0.7.
            \item[h] \textbf{Lagged choice:} this dummy takes value 1 at time $t$ for the route that is chosen at time $t-1$.
            \item[i] \textbf{Choice proportion for a route:} number of times a route is chosen until time $t$/$t$.
            \item[j] \textbf{Total available choice occasions:} total choice occasions used in the choice model (sum of $T$ for all riders).
        \end{tablenotes}
        \end{threeparttable}
}
\end{adjustwidth}
\end{table}

\subsubsection{Choice Model}
We first discuss the results of the compensatory class. In our specification, \textit{Travel Time} implies travel time at the base crowding density (< 1 riders/metre$^2$) and the base standing probability ($< 0.4$). Unlike MNL and LC-MNL, sign and magnitude of parameter estimates of the level-of-service attributes are intuitive in the compensatory class of DLCM (see Table \ref{tab:DLC}). Values of travel time under crowding1-SP1, crowding1-SP2, crowding2-SP1, and crowding2-SP2 are 1.06, 1.09, 1.16, and 1.19 times that of the value in the base condition, respectively. Since only 3\% of riders in our sample have experienced crowding levels above 3 riders/metre$^2$, the value of travel time at crowding density greater than 2 riders/metre$^2$ (crowding2) can be considered as the value of travel time when crowding density is between 2 and 3 riders/metre$^2$. If we further linearly extrapolate the values of travel time under crowding1-SP2 and  crowding2-SP2, metro riders' valuation of travel time appears to increase by around 47\% in extremely crowded condition (crowding levels between 5 and 6 riders/metre$^2$ and standing probability above 0.7) relative to the one obtained under uncrowded conditions (crowding density less than 1 riders/metre$^2$ and standing probability less than 0.4). We also explore the heterogeneity in crowding cost by specifying a lognormal distribution on the marginal disutility of travel time, but the standard deviation does not turn out be statistically significant.   

Since studies based on SP experiments are likely to over-estimate the crowding cost, we compare our crowding cost estimates with those of RP and RP-SP studies \citep[see][for an international comparison]{tirachini2017estimation}. The comparison indicates that the crowding cost estimates of DLCM are lower than the crowding multipliers reported by previous studies. In the RP study of Hong Kong MTR, \cite{horcher2017crowding} find that increase in the disutility of travel time due to an additional rider per square metre is 0.12 times of the base disutility. Similar results are reported based on SP-RP experiments in Santiago \citep{batarce2015use} and Paris \citep{kroes2014value}. 

There are two potential reasons why our results are not directly comparable to those of previous studies. First, we are using a non-random sample of regular riders to estimate the model -- our sample excludes more than 80\% of riders who follow fully non-compensatory choices during our study period (see step 3 in Section \ref{sec:datapro}). These riders might have compensated travel time with crowding when they started using metro, but do not update their route preferences during this experiment. Note that this does not imply that they are insensitive to crowding. Their actual crowding valuation can be lower or higher than our estimates but is not empirically identified due to lack of data on their initial preferences. Whereas earlier studies could identify the crowding cost of this subgroup of riders by making a strong assumption that these riders also make compensatory choices at all occasions, we do not make any such assumption and our crowding valuation estimates are therefore identified using only choice occasions when riders make compensatory choices. Second, the specification of the choice model in DLCM is different as compared to earlier experiments because they control for other trip attributes beside travel time and crowding levels.  

We now discuss the results of the non-compensatory class. Conditional on being in the non-compensatory class, route choice in the last occasion and the historical frequency of chosen routes could explain the future route choices of a rider. Positive signs on coefficients of both covariates indicate that a rider is more likely to choose the route that she has chosen on the last choice occasion and the one that she has chosen more frequently in the past. The non-compensatory class thus captures inertia/habit-based decision rules.    

\subsubsection{Initialisation and Transition Model}

We first discuss the results of the initialisation model. A rider who keeps using the same route and makes fewer transitions over the study period is more likely to be in class 2 (non-compensatory class) at the beginning. These results are aligned with the intuition. We also consider mismatch between the expected and experienced level-of-service, but those covariates do not turn out to be statistically significant.  

In the transition model for both classes, the same covariates are statistically significant. Among the level-of-service attributes, standing probability has a statistically significant association with class transition probabilities. Loosely speaking, positive sign on standing probability indicates that if a rider expects to sit but she has to stand, she is likely to switch to the compensatory class if she is in the non-compensatory class, otherwise is likely to remain in the compensatory class. The proportion of historical transitions at time $t$ and total transitions in the study period also determine the class transition probabilities. Both covariates have different signs and high magnitude. This observation seems counter-intuitive at first because both variables appear to capture similar behaviour. However, it is not and is rather just a consequence of a structural relationship between these covariates. Keeping the proportion of total transitions constant, if a rider has a higher proportion of historical transitions, she is more likely to choose the same route on the next choice occasion, i.e. she is more likely to be in the non-compensatory class. Therefore, a negative sign on the proportion of historical transitions is intuitive and sensible. A positive sign of the proportion of total transitions can be interpreted similarly. 

\subsubsection{Recovering Latent Classes}
We also retrieve the latent class of each rider at different choice occasions using the Viterbi algorithm and then we compute the proportion of all choice occasions when riders are in the compensatory class (class 1). On average, riders follow compensatory decision rules only on 25.5\% of choice occasions. This proportion is 35.1\%, 26.7\%, and 14.4\% for visitors ($T+T_I <10$), less regular commuters ($10 \leq T+T_I \leq 20$), and regular commuters ($T+T_I > 20$), respectively. In essence, frequent riders are more likely to be in the non-compensatory (inertia/habit) state.

\section{Conclusions and Future Work} \label{sec:conc}

The crowding disutility of urban rail users is often computed by eliciting how riders trade travel time for crowding while making a route choice. Previous crowding valuation studies rely on static choice models, overlook learning of riders, and assume that riders always follow a fully compensatory decision rule while ignoring inertia/habit-based choice behaviour. No existing choice model can capture such dynamic semi-compensatory behaviour of riders. 

In this study, we propose a dynamic latent class model (DLCM) which considers heterogeneity in decision rules of a rider by characterising latent classes with decision rules, allows the rider to transition between latent classes over time, and specifies learning behaviour of the rider using instance-based learning theory. We estimate this model by adapting the expectation-maximization (EM) algorithm and recover the most probable sequence of latent classes for each rider using the Viterbi algorithm. We apply the proposed DLCM to estimate the crowding cost of an Asian metro's riders using a two-month-long panel data on their revealed route preferences. 

The empirical data indicate that more than 80\% of riders keep using the same route or shift to another route only once over two months. This observation further strengthens the importance of incorporating inertia/habits in dynamic choice models. The results of DLCM indicate that marginal disutility of travel time of a rider under extremely crowded conditions (crowding density between 5 and 6 riders/meter$^2$ and standing probability above 0.7) is around 47\% higher than the one obtained under uncrowded conditions (crowding density less than 1 riders/meter$^2$ and standing probability less than 0.4). This crowding cost estimate is lower than those reported by previous revealed preference studies. This is perhaps a consequence of an appropriate segmentation of compensatory and inertia-based choice processes. After aggregating the recovered sequence of latent classes for each rider, we find that an average rider follows a compensatory decision rule only on 25.5\% of choice occasions and this proportion further decreases to 14.4\% for regular commuters.  

We note that the crowding cost estimates in this study are obtained from only choice occasions when riders make compensatory choices. However, this raises a natural question of how the consumer benefit of crowding related interventions should be evaluated for those making non-compensatory choices. Intuition suggests that travellers who are making non-compensatory route choices might also perceive the benefits of crowding reduction. However, the welfare calculation approaches for compensatory choices cannot be directly adopted for other non-compensatory decision rules. The subject of benefit calculation under heterogeneous decision rules thus opens up avenues for future research.

We derive the proposed model for two latent classes, but without loss of generality, the model can be extended to multiple classes representing distinct non-compensatory decision rules while accounting for temporal transitions between classes. The potential of DLCM can be explored in understanding dynamic travel behaviour (e.g., preferences for mobility-on-demand services) and various other types of consumer behaviour (e.g., food and healthcare preferences), specially in situations when the longitudinal preference data are easily accessible. 

\newpage 
\bibliographystyle{apalike}
\bibliography{bibliography.bib}

\newpage
\begin{appendices}

\section{Details of the E-step} \label{app:E}

The required expectations in the E-step depend upon $\alpha_{its}(\bm{\Theta})$ and $\beta_{its}(\bm{\Theta})$.  These forward and backward variables are recursively computed as follows: 
\begin{equation}\label{eq:alpha}
    \begin{split}
         \alpha_{its} & =  P(y_{i1}, \dots, y_{it}, q_{its} = 1) \\
         & =  P(y_{i1}, \dots, y_{it} \lvert q_{its} = 1) P(q_{its} = 1) \\
         & = P(y_{i1}, \dots, y_{i(t-1)} \lvert q_{its} = 1) P(y_{it} \lvert y_{i(t-1)}, q_{its} = 1)P(q_{its} = 1) \\
         & = P(y_{it} \lvert y_{i(t-1)}, q_{its} = 1)P(y_{i1}, \dots, y_{i(t-1)}, q_{its} = 1) \\
         & = P(y_{it} \lvert y_{i(t-1)}, q_{its} = 1)\sum_{s^{\prime} = 1}^2 P(y_{i1}, \dots, y_{i(t-1)}, q_{i(t-1)s^{\prime}} = 1, q_{its} = 1)\\
         & = P(y_{it} \lvert y_{i(t-1)}, q_{its} = 1)\sum_{s^{\prime} = 1}^2 P(y_{i1}, \dots, y_{i(t-1)}, q_{i(t-1)s^{\prime}} = 1) P(q_{its} = 1 \lvert q_{i(t-1)s^{\prime}} = 1)\\
         & = P(y_{it} \lvert y_{i(t-1)}, q_{its} = 1)\sum_{s^{\prime} = 1}^2 \alpha_{i(t-1)s^{\prime}} P(q_{its} = 1 \lvert q_{i(t-1)s^{\prime}} = 1)\\
    \end{split}
\end{equation}

We initialise the forward recursion as follows: 
\begin{equation}
    \begin{split}
        \alpha_{i1s} & =  P(y_{i1}, q_{i1s} = 1) \\
        & = P(y_{i1} \lvert q_{i1s} = 1, \text{Inputs}) P(q_{i1s} = 1 \lvert \text{Inputs})
    \end{split}
\end{equation}
\begin{equation} \label{eq:beta}
    \begin{split}
        \beta_{its} & =  P(y_{i(t+1)}, \dots, y_{iT} \lvert y_{it}, q_{its} = 1)\\
        & = \sum_{s^{\prime}=1}^{2} P(y_{i(t+1)}, \dots, y_{iT},  q_{i(t+1)s^{\prime}} = 1 \lvert y_{it}, q_{its} = 1)\\
        & = \sum_{s^{\prime}=1}^{2} P(y_{i(t+1)}, \dots, y_{iT} \lvert y_{it}, q_{its} = 1, q_{i(t+1)s^{\prime}} = 1) P(q_{i(t+1)s^{\prime}} = 1 \lvert q_{its} = 1) \\
        & = \sum_{s^{\prime}=1}^{2} \big[  P(y_{i(t+1)} \lvert  y_{it}, q_{i(t+1)s^{\prime}} = 1)P(y_{i(t+2)}, \dots, y_{iT} \lvert y_{i(t+1)}, q_{i(t+1)s^{\prime}} = 1) \dots \\
         & \quad \dots P(q_{i(t+1)s^{\prime}} = 1 \lvert q_{its} = 1) \big]\\
         & = \sum_{s^{\prime}=1}^{2} P(y_{i(t+1)} \lvert  y_{it}, q_{i(t+1)s^{\prime}} = 1) \beta_{i(t+1)s^{\prime}} P(q_{i(t+1)s^{\prime}} = 1 \lvert q_{its} = 1) \\
    \end{split}
\end{equation}

The backward recursion is initiated by considering $\beta_{iTs} =1 \; \forall i,s$. To validate this recursion, we compute $\beta_{i(T-1)S}$ using equation \ref{eq:beta} and the considered initial condition. 
\begin{equation}
    \begin{split}
     \beta_{i(T-1)S} & = \sum_{s^{\prime}=1}^{2} P(y_{iT} \lvert  y_{i(T-1)}, q_{iTs^{\prime}} = 1) (1) P(q_{iTs^{\prime}} = 1 \lvert q_{i(T-1)s} = 1) \\
         & = P(y_{iT} \lvert  y_{i(T-1)}, q_{i(T-1)s} = 1) 
    \end{split}
\end{equation}

The above equation confirms that $\beta_{iTs} =1 \; \forall i,s$ is appropriate for initialisation of the backward recursion.  

\end{appendices}
\end{document}